\documentclass[acp]{copernicus}

\advance\textheight -1.0cm

\begin{document}

\linenumbers

\title{Generalisation of Levine's prediction
for the distribution of freezing temperatures of droplets:
A general singular model for ice nucleation}

\author{Richard P. Sear}


\affil{Department of Physics, University of Surrey\\ Guildford, Surrey
GU2 7XH, United Kingdom\\
email: r.sear@surrey.ac.uk}

\runningtitle{General singular model for ice nucleation}

\runningauthor{R.~P. Sear}

\correspondence{R.~P. Sear\\ (r.sear@surrey.ac.uk)}

\received{}
\pubdiscuss{} 
\revised{}
\accepted{}
\published{}


\firstpage{1}

\maketitle  

\maketitle

\begin{abstract}
Models without an explicit time dependence, called singular models,
are widely used for fitting the distribution of temperatures
at which water droplets freeze. In 1950 Levine
developed the original singular model. His key assumption
was that each droplet
contained many nucleation sites, and
that freezing occurred due to the nucleation site
with the highest freezing temperature.
The fact that freezing occurs due to the maximum value
out of large number of nucleation temperatures, means
that we can apply the results of what is called extreme-value
statistics. This is the statistics of the extreme, i.e., maximum or
minimum, value of a large number of random variables.
Here we use the results of extreme-value statistics to show that
we can generalise Levine's model to produce
the most general singular model possible.
We show that when a singular model is a good approximation,
the distribution of freezing temperatures should always be given
by what is called the generalised extreme-value distribution.
In addition, we also show
that the distribution of freezing temperatures for droplets of one
size, can be used to make predictions for the scaling of
the median nucleation temperature with droplet size, and vice versa.
\end{abstract}

\introduction

The freezing of water droplets in the Earth's atmosphere is an important
and longstanding
problem \citep{mason_book,pruppacher_book,cantrell05,demott11,sear_rev12}.
We want to understand how the water droplets
freeze, and be able to predict quantitatively the conditions
where the droplets do and do not freeze.
To do this we of course need good experimental data, but
we also need models with few enough parameters
that their values can be reliably obtained by fitting to experimental data.
These models should make as few assumptions as possible, and we should
as clear as possible as to what these assumptions are.
An innovative early
attempt at developing such a model was that of
Levine in 1950 \citep{levine50}.

Levine assumed that water droplets
freeze due to highly variable impurities in the droplets.
He then introduced a simple statistical model of these impurities,
and hence of the freezing behaviour \citep{levine50}.
Levine's model has no direct time dependence. Instead of an explicit rate,
nucleation is assumed to occur at 
a particular nucleation site as soon
as it is cooled to a temperature 
characteristic of that site.
Levine's work has inspired a literature
on what are often
called \citep{pruppacher_book} \lq singular\rq~models
\citep{mason_book,vali08,connolly09,niedermeier10,niedermeier11,murray11,broadley12,welti12}.
By definition singular models are models that lack direct time dependence. 
As far as I know, Levine's is the first such singular model.
He did not call his model singular, that name originates with
\citet{vali66}.
Singular models can be contrasted
with what are called \lq stochastic\rq~models
where there is an explicit nucleation rate for a stochastic 
process of nucleation, and so a direct time dependence \citep{pruppacher_book}.

Levine assumed that each droplet has a large number of
nucleation sites, $N$. He called these sites
\lq motes\rq. He assumed that each mote
had a different 
nucleation temperature, $T_n$, and that the droplet
froze at the highest of these $N$ 
nucleation temperatures.
This second assumption means that, within his model,
the freezing temperature of a droplet, $T_F$, is a 
random number that is the maximum of a number of independent random numbers.
Although Levine apparently did not realise this, this means
that what he was doing was an example of 
what is called extreme-value statistics.
This is the statistics of the extreme (maximum or minimum) of a 
large number of random variables. See the books of either
\citet{jondeau_book}, or \citet{castillo_book}, for an introduction 
to extreme-value statistics.
Incidentally, back in the 1950s,
Turnbull realised that Levine was effectively doing extreme-value statistics \citep{turnbull52_hg}.

Here we use results from modern extreme-value statistics to
show that the results obtained by Levine 
can be written in slightly simpler forms, and 
that they can be generalised -- one of his assumptions was not necessary.
The expression derived by Levine, his Eq.~(2),
is in fact almost (see Appendix A) the probability density 
function of the Gumbel distribution of extreme-value statistics.
If nucleation is indeed occurring on the nucleation site with the highest nucleation temperature, then 
the fraction crystallised should have the form of what is called
the generalised extreme value (GEV) distribution. This is true
for almost all distributions of 
the site nucleation temperatures.
The Gumbel distribution is a special case of the GEV distribution.
Levine also derived a logarithmic dependence of the mode droplet freezing temperature on droplet size.
We will show that this scaling is less general than the Gumbel distribution.

\subsection{Motivation}

Our motivation for this work
is that Levine's key assumptions are very reasonable. These
assumptions are that a droplet has a large 
number of nucleation sites, and that nucleation occurs on the one with the highest nucleation 
temperature.
Also, the neglect of time dependence, although an approximation, simplifies the model, meaning that 
the model has very few parameters.
A model with few parameters is useful, as typically
fitting a model with more than two or three parameters to
experimental data is difficult to justify. The data may not 
adequately constrain the values of a larger number of parameters.

Thus Levine's model seems a very attractive simple model that can be used to fit data directly, and can 
be built on to make more sophisticated models.
For both these reasons it seems worthwhile to use 
results in modern extreme-value statistics to generalise it to produce the most general singular model 
possible, and to determine the minimal assumptions required for it to apply.

In this paper, we will describe Levine's model, and then show how his key results may be
derived using modern extreme-value statistics.
We will then generalise Levine's model
to produce the most general possible singular model of the type that 
Levine introduced. In our final section, we suggest
how this could be used to model experimental data.

\section{Levine's model}
\label{section:levine}

We are interested
in the problem of what happens when a set of nominally identical liquid water droplets are cooled at 
some rate, until they freeze. It is observed 
\citep{levine50,langham58,mason_book,pruppacher_book,niedermeier10,niedermeier11,murray11,vali08,cantrell05,welti12,broadley12}
that the droplets do not all freeze at the same temperature; they freeze over a broad range of 
temperatures.
We want to understand this, and make predictions about this phenomenon, using a simple model.

To do this, we define the probability $P(T_F)$ that a randomly selected droplet has {\em not} frozen, at the 
time when we have cooled it down to a temperature $T_F$. Note that $P(T_F)$ 
is a cumulative probability, 
the probability that a sample freezes between $T_F$ and $T_F-{\rm d}T_F$,
is $({\rm d}P(T_F)/{\rm d}T_F){\rm d}T_F$.
In an experiment, $P(T_F)$ can be approximated by the fraction of a large number of identically 
prepared droplets that are still liquid at a temperature $T_F$.

Levine's model \citep{levine50}
for the freezing of liquid water droplets is simple.
He made the following assumptions:
\begin{enumerate}
\item Each droplet contains 
impurities that have a total of $N$ nucleation sites.
\item Each nucleation site induces nucleation of ice rapidly at a well
defined temperature $T_n$. 
\item This temperature $T_n$ varies from one nucleation site to another.
The sites are 
independent, and the values of $T_n$ are drawn from a probability distribution function 
$p_1(T_n)$.
\item Only one nucleation event is required to induce crystallisation of the droplet, and so the droplet 
crystallises at the highest $T_n$ of its $N$ nucleation sites.
We denote this maximum value of a set of $N$ $T_n$'s, by $T_F$.
\end{enumerate}
Assumption 4 allows us to use extreme-value statistics, see
\citet{castillo_book,jondeau_book} for an introduction to 
these statistics.
Here we define a singular model as being a model in which
assumptions 1 to 4 are made. In particular, assumption 2
eliminates any time dependence, giving us a model with only
a temperature dependence. This definition of a singular model
agrees with that of Pruppacher and Klett \citep{pruppacher_book}.

Levine then made a fifth assumption:
\begin{enumerate}
\setcounter{enumi}{4}
\item The distribution of nucleation temperatures at the sites, $p_1$, is exponential, i.e., 
\begin{equation} p_1(T_n)=s\exp\left(-T_n/w_e\right)/w_e
\label{exp1}
\end{equation}
\end{enumerate}
This distribution has two parameters: $s$ (dimensionless) and $w_e$ (dimensions of temperature). The 
parameter $w_e$ controls how rapidly the probability of finding a site with a given $T_n$, decreases 
with increasing $T_n$. Note that Levine wrote this distribution in a rather
different way. See Appendix A for a comparison to his work that uses notation
that is closer to Levine's.

It is worth noting that
we are only interested in the highest value of $T_n$ of a large number of sites, and so only the 
large $T_n$ tail of the distribution $p_1$ is relevant here.
The form of $p_1$ around average $T_n$ values is 
irrelevant. The maximum is never in this region.
Thus we need only assume that the large $T_n$ tail of the distribution is exponential. The distribution 
around average values can be anything as these sites do not affect freezing and so have no effect in 
experiment. Because of this Eq.~(\ref{exp1}) is only the high-$T$ tail and so is not normalised.
The parameter $s$ controls the location of
this tail, i.e., the bigger $s$ is, the larger the number of 
sites with high nucleation temperatures.

Also, note that we expect $N$ to scale
with the total surface area of the impurities present in a
droplet. So if impurities are deliberately added,
as for example \citet{broadley12} and \citet{welti12} did, then
$N$ should
be proportional to the amount added.
When the impurities are those naturally present in the water, then
if their concentration is constant, their amount
and hence $N$ will be proportional to the droplet volume.

As an aside, we note that
in the language of the
statistical physics of quenched disorder, Levine's model has quenched disorder, 
but no annealed disorder.
The quenched disorder is the variability in the temperatures at which nucleation occurs on the sites. It is 
{\em quenched} disorder as it is assumed not to depend on time,
but to be fixed for a given droplet.
There is no annealed disorder as there is no time dependence.
Annealed disorder is associated with dynamic fluctuations,
which are neglected in singular models.

\section{Modern derivation of Levine's key results}

If we make all 5 assumptions of section \ref{section:levine},
we can easily derive the Gumbel distribution for the
freezing temperature, $T_F$.
The derivation of Levine's
distribution of freezing temperatures proceeds as follows.
We start by obtaining the cumulative probability distribution function
for a nucleation site, $P_1$. $P_1$ is the probability that 
the nucleation temperature at a nucleation site is {\it lower} than $T_n$.
The cumulative probability $P_1$ is just a definite integral over $p_1$,
so using the exponential $p_1$ of Eq.~(\ref{exp1}), we have
\begin{eqnarray} 
P_{1}(T_n)&=&1-\int_{T_n}^{0}p_1(T){\rm d}T
\simeq 1-\int_{T_n}^{\infty}p_1(T){\rm d}T  \\
&\simeq &1-s\exp\left(-T_n/w_e\right)
\end{eqnarray}
where we used Eq.~(\ref{exp1}) for $p_1$,
and we extended the upper limit on integration from $0^{\circ}$C to infinity.
Of course, $p_1$ must be zero for temperatures above $0^{\circ}$C, and so the
approximate $p_1$ of Eq.~(\ref{exp1}) is only valid for values of $s$ and $w_e$ such that
the exponential $p_1$ is negligible
for $T_n\ge 0^{\circ}$C. We assume this to be the
case here.

If a droplet contains $N$ nucleation sites then the probability that
it is in the liquid phase, $P(T_F)$, is 
simply the probability that
{\em all} $N$ nucleation sites have crystallisation temperatures below $T_F$.
We are assuming that even a single nucleation site will cause
freezing.
As 
these nucleation sites are independent
this probability is just $P_{1}^N$, so,
\begin{eqnarray} 
P(T_F,N)&=& P_{1}^N =\left[1-s\exp\left(-T_F/w_e\right)\right]^N
\\
&\simeq & \exp\left[-Ns\exp\left(-T_F/w_e\right)\right]
\end{eqnarray}
Here we used
the fact that when $N$ is large, we are interested in the range when $s\exp\left(-T/w_e\right)\ll 1$, and 
so we can use the result $(1+x)^n\simeq \exp(nx)$, which is
valid for small $x$ and large $n$.

\begin{figure}[tb]
\begin{center}
\includegraphics*[width=6.5cm]{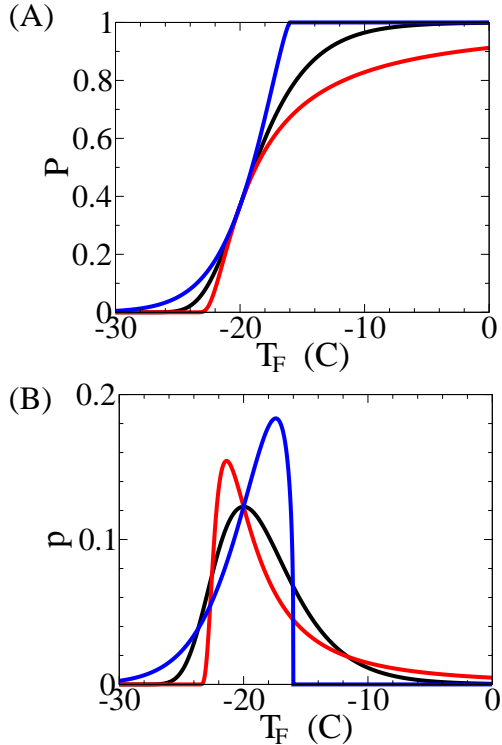}
\caption{Plots of: (A) the cumulative
distribution function $P(T_F)$, and (B) the
probability density function $p(T_F)$. $T_F$ is
the temperature a droplet freezes.
In both plots the black, red and blue curves
are the Gumbel, Fr\'{e}chet
and Weibull distribution functions, respectively.
For the Fr\'{e}chet distributions $\xi=0.75$, while for 
the Weibull distributions $\xi=-0.75$.
For all curves, the location 
$\mu=-20^{\circ}$C, and the width $w=3^{\circ}$C.
For the $P$'s we
use Eq.~(\ref{gen_exp}).
Note that
for the Gumbel distribution, $\mu$ and $w$ are related
to $s$, $N$ and $w_e$, by $\mu=w_e\ln(sN)$ and $w=w_e$.
\label{gev}}
\end{center}
\end{figure}

We can rewrite $P$ as
\begin{equation}
P(T_F,N)= \exp\left[
-\exp\left(-\left[T_F-w_e\ln(Ns)\right]/w_e\right)
\right]
\label{gumbel_exp}
\end{equation}
This is the cumulative distribution function for the Gumbel extreme-value distribution 
\citep{castillo_book,nicodemi_chapter,jondeau_book}.
The Gumbel distribution is a special case of the GEV distribution.
This Gumbel $P$ is plotted in Fig.~\ref{gev}(A).

\begin{figure}[tb]
\begin{center}
\includegraphics*[width=7.0cm]{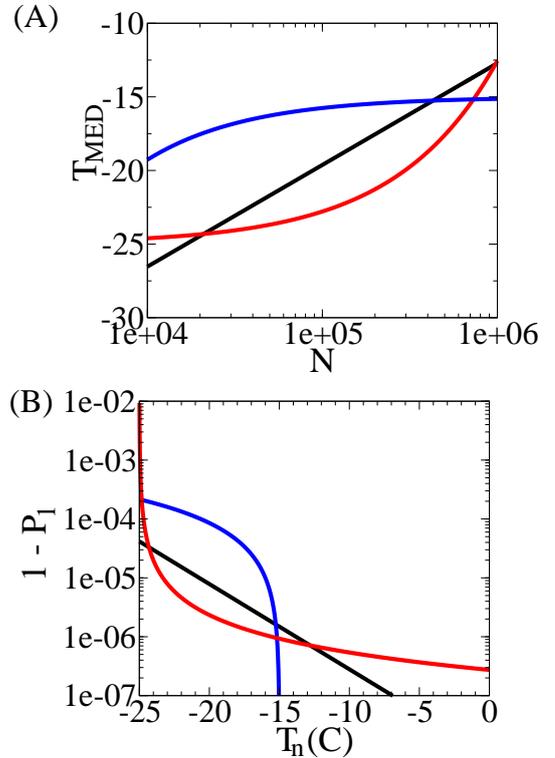}
 \caption{(A) Plot of the scaling
of the median freezing temperature, $T_{MED}$,
with the number of nucleation sites, $N$.
The black curve is the $\ln N$ scaling that
is consistent with the Gumbel distribution.
It is a plot of the $T_{MED}$ of Eq.~(\ref{tmed_gumbel}),
with parameters $s=10^{-8}$ and $w_e=3^{\circ}$C.
The red curve is the $N^{3/4}$ scaling that
is consistent with the Fr\'{e}chet distribution with $\xi=0.75$.
It is a plot of the $T_{MED}$ of Eq.~(\ref{tmed_frechet}),
with parameters $b=2\times 10^{-5}$ and $T_L=-25^{\circ}$C.
The blue curve is the $N^{-3/4}$ scaling that
is consistent with the Weibull distribution with $\xi=-0.75$.
It is a plot of the $T_{MED}$ of Eq.~(\ref{tmed_weibull}),
with parameters $c=10^{-5}$ and $T_U=-15^{\circ}$C.
(B) Plot of cumulative distribution
functions for the nucleation temperatures, $T_n$, at the nucleation
sites. We clarity, we plot $1-P_1(T_n)$, i.e., the probability
that a nucleation site has a nucleation temperature {\em above} $T_n$.
This tends to 0 not 1 at high $T$.
The black, red and blue curves are $1-P_1$'s that yield
Gumbel, Fr\'{e}chet, and Weibull
distributions, respectively. Values of the parameters are the
same as in (A).
\label{n_scale}}
\end{center}
\end{figure}

For an exponential $p_1$,
the width of the Gumbel distribution of crystallisation temperatures is 
the same as the width $w_e$ for a single nucleation site.
The median crystallisation temperature, $T_{MED}$,
is obtained by noting that, by the definition
of the median, $P(T_{MED})=1/2$. Then we have that
the median freezing temperature is
\begin{equation} 
T_{MED}(N)=w_e\ln s-w_e\ln(\ln 2)+w_e\ln(N)
\label{tmed_gumbel}
\end{equation}
The scaling of the average freezing 
temperature with the number of nucleation sites is logarithmic,
as Levine found in his Eq.~(10). The variation of $T_{MED}$
with $N$ is shown in Fig.~\ref{n_scale}(A).
Finally, the probability density function for nucleation
to occur at a temperature $T_F$, $p(T_F)$, is 
\begin{eqnarray} 
p(T_F)&=&\frac{{\rm d}P(T_F)}{{\rm d}T_F}
\nonumber\\
&=&
P(T_F)\exp\left[-\left(T_F-
w_e\ln(Ns)\right)/w_e\right]/w_e
\end{eqnarray}
This is almost equal to Levine's Eq.~(2) for $p(T)$.
It is not quite equal as Levine made a small approximation.
We compare our expressions with Levine in 
detail in Appendix A.
The Gumbel $p$ is plotted in Fig.~\ref{gev}(B). Note that
the Gumbel distribution has a characteristically fatter tail on
the high-temperature side than on the low-temperature side
of its maximum. This is often seen in experimental data
for the freezing of water droplets, for example in Fig.~52
of \citet{langham58}.

\section{Predictions of modern extreme value statistics}

The Gumbel distribution that Levine derived
is just one of the three types or classes
of extreme-value distributions,
that together make up
the GEV \citep{nicodemi_chapter,jondeau_book,castillo_book}.
The other two are the Weibull and Fr\'{e}chet distributions.
In brief, modern extreme value theory allows us to show that
for any singular model,
$P(T_F)$ should be given by the GEV. The requirements
that must be satisfied are only that
assumptions 1 to 4 must hold, and that $p_1$ should be
a simple continuous function of $T$ in the temperature
range of interest.

In this section we use modern extreme-value statistics to generalise
Levine's findings, in order to obtain the more general
GEV form of $P(T_F)$. We also derive the scaling
of the average freezing temperature, $T_{MED}$,
with $N$ for the Weibull and Fr\'{e}chet distributions,
and compare the results with experimental data.
We will briefly consider how good our assumption
that $N$ is the same in droplets of the same volume.
Also, as our model is in the singular limit,
we outline a criterion for the singular limit to be a good approximation.

\subsection{The generalised extreme value distribution}

Once assumptions 1 to 4 are made, the freezing temperature
$T_F$ is the maximum of a large number of independent
identically distributed random variables. Then it can be shown
that the cumulative distribution function for $P(T_F)$
is given by the GEV. This is true under only weak
conditions on $p_1$ \citep{nicodemi_chapter,castillo_book,jondeau_book},
The GEV is conventionally written as
\citep{castillo_book,nicodemi_chapter,jondeau_book}
\begin{equation}
P(T_F)=\left\{
\begin{array}{lc}
 \exp\left[
-\exp\left(-\left(T_F-\mu\right)/w\right)
\right] & \xi=0 \\
 \exp\left[
-\left(1+\xi\left(T_F-\mu\right)/w\right)^{-1/\xi}
\right] & \xi\ne 0
\end{array} \right.
\label{gen_exp}
\end{equation}
This is a three-parameter cumulative probability
distribution function. Assumption 5 is not required to derive it.
The parameters are a width
parameter $w$, a location parameter $\mu$, and an exponent
$\xi$. The value of the parameter $\xi$ controls the class
of the GEV.
With $\xi=0$, the GEV is the Gumbel distribution, while
for $\xi>0$, the GEV is the Fr\'{e}chet distribution, and
for $\xi<0$, it is the Weibull distribution.
Examples of all three distributions are plotted in Fig.~\ref{gev}.


Equation (\ref{gen_exp}) generalises the Gumbel distribution,
Eq.~(\ref{gumbel_exp}), that Levine derived.
Whereas the Gumbel distribution is produced by exponentially decaying
(in the sense of decaying faster than any power law)
$p_1$'s,
almost any continuous simple $p_1$ will lead
to the GEV. This includes $p_1$'s that decay as power laws. Power
law $p_1$'s
lead to the Fr\'{e}chet limit of the GEV, and $p_1$'s
with upper limits lead to the Weibull distribution.

The form of the distribution of nucleation temperatures, $p_1$,
also determines the scaling of the median freezing temperature
with $N$. We have already seen
that for an exponential $p_1$, this scaling is $\ln N$,
Eq.~(\ref{tmed_gumbel}).
This $p_1$ also leads to a Gumbel $P(T_F)$, but other $p_1$'s
lead to the same Gumbel form for $P$ but have different
scaling of $T_{MED}$ with $N$.
For example a Gaussian $p_1$ leads to a $(\ln N)^{1/2}$ scaling
\citep{castillo_book}.

What this means is that if, for example, data is well fit
by a Gumbel $P$, i.e., $\xi\simeq 0$, then we cannot argue that
$T_{MED}$ scales as $\ln N$ -- although it should be noted
that $\ln N$ and $(\ln N)^{1/2}$ scaling are relatively
similar so if it is a Gumbel then we do have a rough idea
of the scaling of $T_{MED}$.
However, if data is clearly best fit by a $\ln N$ scaling
of $T_{MED}$, then this is good evidence that $p_1$ is indeed
an exponential function of $T_n$, in the temperature range
of interest. It is stronger evidence for an exponential $p_1$,
than the Gumbel distribution providing a good fit to $P$.

Fitting the GEV could be done following the same methods used to
fit the GEV to data in other fields. The book of \citet{castillo_book} on extreme value statistics
discusses general fitting approaches. It is worth noting that he does not recommend the
standard unweighted least-squares fitting procedure as that gives a low weight to errors in
the tail of $P(T_F)$. See \citet{castillo_book} for suggested weighting functions to be minimised
in fitting. \citet{castillo_book}
also discusses the fact that plots of
$\ln[\ln(1/P(T_F))]$ as a function of $T_F$, show a characteristic curvature that depends on $\xi$.
This
can be used to differentiate between Gumbel, Fr\'{e}chet and Weibull distributions.
Such a plot should be a straight line if the data follows the Gumbel distribution,
while it will curve down for Weibull-distributed data, and up for Fr\'{e}chet-distributed data.

\cite{jondeau_book} discuss a related method, which uses what are called
Quantile-Quantile or Q-Q plots. Here the temperature at which the GEV
function for $P$ is a particular value, is plotted as a function of
the temperature in the data which gives the same value for $P$.
When this is done, then if the data is indeed well approximated by the GEV,
and the correct value of $\xi$ is chosen, then the Q-Q plot will be
a straight line (arbitrary values of $\mu$ and $w$ can be used
as they just change the slope and intercept of the plot).
See
\cite{jondeau_book} for details. They
also consider the application of maximum likelihood methods to obtaining
the most reliable estimates of $\xi$, $\mu$ and $w$.

In the next section, we outline how both the
Fr\'{e}chet and Weibull distributions can be derived from their
respective $P_1$'s. This also allows us
to also determine how the median freezing temperature,
$T_{MED}$, scales with $N$.

\subsection{Brief derivation of the Fr\'{e}chet and Weibull distributions}
\label{f&w}

In this section we briefly show how the 
Fr\'{e}chet and Weibull distributions can be derived from the
$P_1$'s of the nucleation sites, where as before
$P_1(T_n)$ is the probability that the nucleation temperature at a site
is below $T_n$.
As the $N$ nucleation sites are independent, we always
have that the probability that a droplet has not frozen
at a temperature $T_F$ is
\begin{equation}
P(T_F)=P_1^N(T_n)
\end{equation}
which as we are in $N\gg 1$, and $1-P_1\ll1$ limit can be written as
\begin{eqnarray}
P(T_F) &=&\left[1-[1-P_1(T_F)]\right]^N(T_n)
\nonumber\\
&\simeq&
\exp\left[-N[1-P_1(T_F)]\right]
\label{ptf_gen}
\end{eqnarray}
Armed with this relation, we start with the
Fr\'{e}chet distribution.
The Fr\'{e}chet distribution results from a power-law
cumulative distribution, $P_1$, for nucleation temperatures
$T_n$,
\begin{equation}
P_1(T_n)=1-\frac{b}{(T_n-T_L)^{1/\xi}}~~~~~~\xi>0
\label{p1_frechet}
\end{equation}
Note that this is a power-law decay with a lower cutoff,
$T_L$. The parameter $b$ (like $s$)
controls the size of the tail. This expression holds for the
large $T_n$ tail, where $P_1$ is close to 1.
Note that here, we have the restriction that $\xi>0$,
so this is a power-law decay of $p_1$ with $T_n$.
In Fig.~\ref{n_scale}(B), we have plotted an example $P_1$.
We plot $1-P_1$ not $P_1$ itself, as $1-P_1$ decays to 0, and
this is a little clearer to see than a decay to 1.
The cumulative probability $1-P_1(T_n)$ is the probability
that a nucleation site has a nucleation temperature above $T_n$.
Now, using the $P_1$ of Eq.~(\ref{p1_frechet}) in
Eq.~(\ref{ptf_gen}), we have the Fr\'{e}chet distribution
\begin{equation}
P(T_F)\simeq
 \exp\left[-\frac{N b}{(T_F-T_L)^{1/\xi}}\right]
\label{ptf_frechet}
\end{equation}
Note that $N$ and $b$ always appear as their product, $Nb$.
Therefore, the freezing behaviour does not depend on $N$ and
$b$ separately, only on their product.

We now consider the Weibull distribution.
The Weibull distribution results from a 
cumulative distribution, $P_1$, with an upper cut off
\begin{equation}
P_1(T_n)=\left\{\begin{array}{cc}
1-c(T_U-T_n)^{-1/\xi} & T_n\le T_U \\
1& T_n > T_U
\end{array}\right.
~~~~~~\xi<0
\label{p1_weibull}
\end{equation}
where
$T_U$ is the upper cutoff, and $c$ is a parameter that (like $s$ and $b$)
controls the size of the tail. This expression holds for the
large $T$ tail, where $P_1$ is close to 1.
Note that here, we have the restriction that $\xi<0$,
so $1-P_1$ is a positive-exponent power-law
function of $T_n$. An example $1-P_1$ is plotted
in Fig.~\ref{n_scale}(B).

In the singular limit, a hard upper cutoff, $T_U$, to the distribution
of nucleation temperatures, is possible. In experiment,
there will presumably be a limit to how well defined this cutoff temperature
can be. In practice, the Weibull model should be a good model for experimental data when the inevitable uncertainty in $T_U$, call it $\delta T_U$, is much smaller than the range of temperatures over which nucleation occurs. This range of temperatures could be measured by the standard deviation of the observed nucleation temperatures, $\sigma_F$. So when
$\delta T_U \ll \sigma_F$, and the Weibull model fits the data well, the Weibull model should be useful.

Returning to the $P_1$ of Eq.~(\ref{p1_weibull}). If we put this in
Eq.~(\ref{ptf_gen}), we have the Weibull distribution
\begin{equation}
P(T_F)\simeq
 \exp\left[- N c(T_U-T_F)^{-1/\xi}\right]
\label{ptf_weibull}
\end{equation}

Having derived the
Gumbel, Fr\'{e}chet, and Weibull distributions, we can compare them.
Example plots are shown in Fig.~\ref{gev}.
The differences between the three distributions
is particularly clear in the plots of their probability
densities in Fig.~\ref{gev}(B).
The Fr\'{e}chet distribution has a much fatter high-temperature
tail than the Gumbel, and a low-temperature cutoff.
So, if the GEV is fit to data with such a sharp lower-temperature
cutoff and/or fat tail, the best fit may be with a $\xi>0$,
implying that a Fr\'{e}chet distribution is a better model than
a Gumbel. The fatter tail of the Fr\'{e}chet comes
from a power-law tail in $p_1$, i.e., from a fatter tail
in the distribution in the nucleation temperatures at
the individual sites.
By contrast,
the Weibull distribution has a high-temperature cutoff, which
implies a high-temperature cutoff in $p_1$. For
data with a sharp upper cutoff to nucleation, the Weibull
model may be best.

\subsection{Scaling of $T_{MED}$ with droplet volume and
surface area of added impurity}

An exponential $P_1$ led to a Gumbel distribution,
and $\ln N$ scaling
of the median freezing temperature with $N$. Here
we derive the corresponding scalings with system size for
power-law $P_1$'s, and $P_1$'s with upper limits.

Power-law $P_1$'s lead to the Fr\'{e}chet $P(T_F)$
of Eq.~(\ref{ptf_frechet}).
The median freezing temperature, $T_{MED}$, is the temperature
at which $P=1/2$, and so here we have
\begin{equation}
T_{MED}=T_L+\left(\frac{b}{\ln 2}\right)^{\xi}N^{\xi}
\label{tmed_frechet}
\end{equation}
The median freezing temperature is a power-law function of the
number of nucleation sites, $N$.
This is illustrated in Fig.~\ref{n_scale}(A).

$P_1$'s with an upper cutoff lead to the Weibull $P(T_F)$
of Eq.~(\ref{ptf_weibull}).
The median freezing temperature, $T_{MED}$, is again the temperature
at which $P=1/2$, and so here we have that
\begin{equation}
T_{MED}=T_U-\left(\frac{c}{\ln 2}\right)^{\xi}N^{\xi}
\label{tmed_weibull}
\end{equation}
The median freezing temperature approaches the upper limit,
$T_U$, of the nucleation temperatures, as $N\to\infty$.
This is shown in Fig.~\ref{n_scale}(A).
This hard cutoff to the nucleation temperatures will presumably
be only an approximation to the truth. However,
Eq.~(\ref{tmed_weibull}) should be a good approximation when
the inevitable uncertainty in $T_U$ is small in comparison
with the change in $T_{MED}$ with $N$.

Having determined the scaling of $T_{MED}$ with $N$ for
all three classes of the GEV, we can compare these predictions
with experimental findings.
There have been a number of studies of the average freezing
temperature of droplets. Both the droplet volume, and the
surface area of added impurity have been varied.
A plot of the average nucleation temperatures obtained in early
work is shown in Mason's book \citep{mason_book}, in Fig.~4.2.
On the log-linear scale, some data is linear, which is
consistent with an exponential-tailed $p_1$, whereas other
data sets appear to be plateauing at large droplets, suggesting
an upper cutoff to $p_1$.

In more recent work,
both \citet{broadley12}, and
\citet{welti12} have studied average
nucleation temperatures as a function of the surface area
of added clay particles. The clay is illite for
Broadley {\it et al.}, and kaolinite for
Welti {\it et al.}.
We expect the number of nucleation sites, $N$, to scale
with the surface area of added clay.
\citet{broadley12}'s data
seem to be plateauing at large amounts of added illite clay.
This is in their Fig.~4.
\citet{welti12} observe a logarithmic scaling
of the median nucleation temperature with clay surface area.
Thus, the data on the scaling
of the freezing temperature with system size, suggests that
ice nucleation is occurring on
sites with either an exponentially decaying $p_1$, or a $p_1$ with an
upper cutoff.

\subsection{Validity of the asssumption that $N$ is the same for all
droplets}

If in experiment, the variable is the amount of an impurity that is added,
then  it seems a safe
assumption that $N\propto$~surface area of added impurity, and that two
droplets with the same amount of added impurity have the same number
of nucleation sites, $N$.
This just relies on there being a density of nucleation sites on the surface,
that is approximately constant.

However, if the variable is droplet volume $V$, then we are relying on
$N\propto V$ and each droplet having the same number of nucleation sites, $N$.
If the nucleation sites are distributed over a large number, $n$, of impurity
particles then the Central Limit Theorem of statistics tells us
that the variation in $N$ from one droplet to another will be of order
$N/n^{1/2}$. Thus for $n\gg 1$ this will be small and our assumption
of constant $N$ will be only a small approximation. However,
if $n$ is small, i.e., each droplet has only a few impurity particles, then even
though each may have many nucleation sites, there will be large fluctuations in
$N$ from one droplet to another of the same volume.
These fluctuations could potentially cause deviations from the (GEV) distribution, due to some droplets having many more nucleation sites than others.

\subsection{Validity of the singular limit}

The assumption that nucleation occurs at a site at a
precisely determined temperature,
$T_n$, is presumably only an approximation
to the truth. If ice nucleation
in a droplet occurs at a temperature-dependent stochastic
rate, $R(T)$,
then nucleation will
occur over a temperature range of some width $\Delta T_S$.
This width is expected to scale as
\begin{equation}
\Delta T_S=\left(
\frac{1}{R}\frac{\partial R}{\partial T}\right)_{R=R_{COOL}}^{-1}
\end{equation}
The expression in brackets is the ratio of the temperature derivative of
the rate, to the rate itself. One over this ratio is an approximation to
the change in temperature needed to double the nucleation rate.
This ratio is evaluated at a temperature such that the nucleation rate, $R$,
equals the cooling rate, $R_{COOL}$, in experiment.
Note that it is non-negligible assumption that a well-defined
nucleation rate exists in these systems \citep{sear13}.

In words, the expected spread in nucleation temperatures, $\Delta T_S$ due
to a temperature-dependent nucleation rate,
is approximately equal to the temperature change
needed to double the nucleation rate.
This temperature change is evaluated when the nucleation rate equals
the cooling rate.

The singular limit is then the limit $w\gg\Delta T_S$.
When the width in the spread of freezing temperatures due
to the spread in characteristic nucleation temperatures, $T_n$,
is much larger than the spread due to the stochastic nucleation rate,
then singular models can be a good approximation to experimental data.
But when the spread due to the stochastic nature of the nucleation,
$\Delta T_S$ is comparable to that due to the variability in nucleation
temperatures, then singular models will be poor approximations.

\conclusions

Singular models have been and are being used to fit experimental
data \citep{mason_book,pruppacher_book,vali08,niedermeier10,broadley12}.
The fact that they work so well suggests
that in many situations an explicit time dependence 
does not need to be considered.
Here we have shown within a general singular model, the distribution of
freezing temperatures should be given by the GEV.
This follows
if, as \citet{pruppacher_book} do,
a singular model is defined as being assumptions 1 to 4, and
$p_1$ is a simple function of temperature.

There is a caveat to this statement. This is that for $P(T_F)$
to be given by the GEV, it is necessary that over the
temperature range of interest, $P_1(T_n)$ should be given
by a single continuous function, such as a power law or
exponential. This may not be the case if there is more
than one type of nucleation site (perhaps due to multiple
particle species) which all make significant contributions
to $P_1$ but have different dependences on temperature. Thus it
may be that even in the singular limit, $P(T_F)$
deviates from the GEV in the presence of nucleation
on a complex mixture of impurities. Then there is no
general theory. Here calculating $P(T_F)$ can only be done
if the distribution of nucleation temperatures at the sites
is known $p_1$. This will presumably be difficult even for
simple impurities. However, if we have experimental data
for $P(T_F)$, then Eq.~(\ref{ptf_gen}) tells us that
if we plot $\ln P(T_F)$ as a function of $T_F$, then
we should be plotting $-N[1-P_1]$. Then what we are plotting
is directly proportional to the cumulative probability
of finding a nucleation site with a nucleation temperature
above $T_F$. This may aid in interpreting data for $P(T_F)$.

Microscopic models of nucleation, for example
those based on classical nucleation theory,
are also used to fit and understand experimental results
\citep{cantrell05,niedermeier11}. They
can provide insight into droplet freezing data
that a purely statistical model such as an extreme-value-statistics
model cannot provide. However, 
in the singular limit ($\Delta T_S\ll w$)
almost any microscopic
model will give the GEV. Thus in this limit any two
microscopic models with similar $P_1$ will be
essentially equivalent.

Finally, in practice if data deviates from the GEV, it may be difficult
to assess why, as there could be several reasons for the deviations.
These include: 1) effects of a stochastic temperature-dependent rate,
of the type that classical nucleation theory predicts; 2) a complex $p_1$
due to a mixture of surfaces, all making significant contributions
to nucleation; 3) non-classical-nucleation-theory time-dependent
processes, for example, irreversible chemical processes at
surfaces that change the ability of a surface to promote ice
nucleation; (4) each droplet contains only a handful of impurity
particles with the nucleation sites, and so some droplets have many
more nucleation sites ($N$) than others.
Distinguishing between the four may be difficult, although varying the cooling rate may be one way to eliminate at least some of them.

\subsection{Suggestions for future work}

It may be worthwhile to do what is standard practice
in other fields where extreme-value statistics are used,
and to fit the GEV distribution to the data.
Here the data is the fraction of droplets
that have frozen, as a function of temperature.
If the fit is good, then the data would be
consistent with an extreme-value model, and if
the fitted $\xi$ is close to zero, it would suggest
that the high $T$ tail of the nucleation temperatures
of individual sites is indeed exponential or similar, i.e.,
decays faster than a power law \citep{nicodemi_chapter}.
However, a value of $\xi>0$ suggests a power law decay
for $p_1$, while $\xi<0$ suggests an upper limit
beyond which $p_1=0$. In other words, the value of $\xi$
gives information on the form of $p_1$.

Another point of view, is that assumptions 1 to 4 (only), lead
to the GEV, and so the GEV can be used to decouple
assumptions 1 to 4, from assumption 5.
Assumptions 1 to 4 are presumably only approximately true.
In particular, assumption 2 that a site induces nucleation
at a temperature independent of cooling rate is presumably
only approximate. To rigorously test for violations of
this assumption, which is at the heart of
singular models, we would like to avoid assumption 5, and
so should tests for deviations from the GEV, not from the Gumbel
distribution.

A final point to note is that the
high-$T$ tail in $p_1$, not only determines $P(t)$,
but also determines the scaling of the median nucleation
temperature with $N$. 
In general, the fatter the tail
in $p_1$, the faster the median nucleation temperature varies
with $N$. This is illustrated in Fig.~\ref{n_scale}(A).
So if a fit to a $P(t)$ produces a $\xi>0$ then the
volume dependence should be faster than logarithmic, the
median freezing temperature should scale as $N^{\xi}$.
A best fit value of $\xi<0$ suggests a Weibull distribution,
which has an upper cutoff and hence an upper limit to
the median nucleation temperature as droplet volume
is increased.

\appendix
\section{Comparison with Levine's expression}

Levine's approximation for the probability that nucleation has
not occurred at a temperature $T$ is the first factor in his
Eq.~(2). We write this as
\begin{equation}
P(T)=\left(1-\frac{1}{\mu}\right)^{ar^{-T}}~~~~r>1
\label{levine1}
\end{equation}
where we have taken the dominant term in his exponent, $ar^{-T}$,
and changed what is a $+T$ in Levine's expression to a $-T$.
Levine uses the absolute value of $T$ in Celsius, so his
$T$ is our $-T$. In this expression $\mu=V_R/\Delta V$, where
$\Delta V$ is the volume of a droplet, and
$V_R$ is a large reservoir volume, $\gg \Delta V$.
The droplet volume $\Delta V$ is proportional
to our $N$. The parameter $a$ is
is analogous to our $s$ parameter.
The $r$ parameter controls the width of Levine's
distribution, so it is analogous to our $w_e$.

If we note that both $\mu$ and $ar^{-T}\gg 1$, we can rewrite 
Eq.~(\ref{levine1}) as an exponential
\begin{eqnarray}
P(T,\mu)&=&\exp\left[-\frac{ar^{-T}}{\mu}\right]
\nonumber\\
&=&\exp\left[-\exp\left[-T\ln r+\ln\left(\frac{a}{\mu}\right)\right]\right]
\label{levine2}
\end{eqnarray}
If we compare this equation with Eq.~(\ref{gumbel_exp}),
we see that they are the same if $\ln r=1/w_e$,
and $a/\mu=Ns$. Also, from this equation it
is easy to show that the median nucleation temperature,
$T_{MED}$, scales as $\ln(1/\mu)\propto\ln\Delta V$.

Levine's Eq.~(2) is actually his approximation for
the probability density, $p$, that nucleation has occurred at a temperature $T$
not the cumulative probability that it has not occurred
down to a temperature $T$. This $p={\rm d}P/{\rm d}T$.
The expression in Levine's Eq.~2 is not quite the $T$ derivative
of Eq.~(\ref{levine2}), as Levine treats $T$ as a discrete
variable when it is a continuous variable. Thus the expression in
his Eq.~(2)
is, for this reason, approximate.
But this should not obscure the fact that Levine was the first
to realise that the extremes of the distribution of nucleation sites
determine the nucleation behaviour, and that the use of what is
essentially extreme-value statistics can be used to model
freezing behaviour.

\begin{acknowledgements}
It is a pleasure to thank James Mithen for bringing Turnbull's and
hence Levine's work to my attention, and Ray Shaw for helpful
discussions.
I acknowledge financial support from EPSRC (EP/J006106/1).
\end{acknowledgements}


\end{document}